\documentclass[11pt, peereview, onecolumn, letterpaper]{IEEEtran}

\usepackage{amsmath}
\usepackage{amssymb}
\usepackage{amsmath}
\usepackage{amsthm}
\usepackage{amssymb}
\usepackage{amscd}
\usepackage{latexsym}
\usepackage{multirow}
\usepackage{graphics}
\usepackage{color}
\usepackage{cite}
\usepackage{mathtools}
\usepackage{amsfonts}
\usepackage[font=footnotesize]{caption}
\usepackage{setspace}

 \usepackage{caption}
\usepackage{subcaption}

\date{}

\newtheoremstyle{mynewtheorem}
{5pt}
{5pt}
{\it}
{}
{\bf}
{.}
{.5em}
{\thmname{#1}\thmnumber{#2}}%

\theoremstyle{mynewtheorem}
\newtheorem{myprop}{Proposition }
\newtheorem{mylemma}{Lemma }

\newtheorem{mycol}{Corollary }

\usepackage{hyperref}

\newcommand{\m}[1]{\mathbf{#1}}
\newcommand{\tr}[1]{\mathrm{Tr}\left\{#1\right\}}
\newcommand{\s}[1]{\mathrm{sgn}\left(#1\right)}
\newcommand{\unt}[1]{\mathrm{unt}\left(#1\right)}
\newcommand{\opt}[1]{\m #1^{\text{\normalfont opt}}}

\newcommand{\mom}{\boldsymbol\omega}
\newcommand{\mphi}{\boldsymbol\phi}
\newcommand{\Diag}{\mathrm{Diag}}

\newcommand{\argmax}[2]{\underset{#1}{\mathrm{argmax}}~{#2}}
\newcommand{\argmin}[2]{\underset{#1}{\mathrm{argmin}}~{#2}}

 \usepackage{graphicx}
 \usepackage{subcaption}
 \DeclareCaptionLabelFormat{r-parens}{{#2}} 
\renewcommand{\thesubsection}{\Roman{section}.\Alph{subsection}}

\begin{document}

\title{\hfill{\small{\it Submitted to IEEE Transactions on Signal Processing}}\\
\vspace{1cm}\renewcommand{\baselinestretch}{1}\Huge L1-norm Principal-Component Analysis of\\ Complex Data}
 
\author{
		\thanks{Some preliminary studies related to this paper were presented at the IEEE International  Workshop on Signal Processing Advances in Wireless Communications (SPAWC), Stockholm, Sweden, in June 2015 \cite{L1PCA_28}.
	 This work was supported in part by the National Science Foundation under Grant ECCS-1462341 and  the Office of the Vice President for Research of the Rochester Institute of Technology.}
Nicholas Tsagkarakis$^{\ddag}$,
Panos P. Markopoulos$^\dag$,
Dimitris A. Pados$^{\ddag *}$
\footnote{$^*$Corresponding author.}

\IEEEauthorblockA{
	\vspace{0.2cm}
	\vspace{-0.2 cm} $^\ddag$Department of Electrical Engineering\\
	\vspace{-0.3 cm} University at Buffalo, The State University of New York\\
	\vspace{-0.3 cm} Buffalo, NY 14260 USA\\
	\vspace{-0.3 cm} E-mail: \texttt{\{ntsagkar, pados\}@buffalo.edu} \\ 	
\vspace{0.2cm}
\vspace{-0.3 cm} $^\dag$Department of Electrical and Microelectronic Engineering \\
\vspace{-0.3 cm} Rochester Institute of Technology\\
\vspace{-0.3 cm} Rochester, NY 14623 USA\\ 
\vspace{-0.3 cm} E-mail: \texttt{panos@rit.edu} \\
 }
EDICS: MLR-ICAN, MLR-LEAR, MLR-PATT, MDS-ALGO, SSP-SSAN \\
 \vspace{-0.1cm}
 Submitted: \today
}

\setlength{\textheight}{9in}
\pagestyle{plain}

\maketitle

\thispagestyle{empty}

\begin{abstract}
L1-norm Principal-Component Analysis (L1-PCA) of real-valued data has attracted significant research
interest over the past decade. However, L1-PCA of complex-valued data remains to date unexplored despite the
many possible applications (e.g., in communication systems). In this work, we establish theoretical and algorithmic
foundations of L1-PCA of complex-valued data matrices. Specifically, we first show that, in contrast to the real-valued
case for which an optimal polynomial-cost algorithm was recently reported by Markopoulos et al., complex L1-PCA
is formally NP-hard in the number of data points. Then, casting complex L1-PCA as a unimodular optimization
problem, we present the first two suboptimal algorithms in the literature for its solution. Our experimental studies illustrate the sturdy resistance of complex L1-PCA against faulty measurements/outliers in the processed data.
\end{abstract}

 \vspace{-0.3cm}

\section{Introduction}
For more than a century, Principal-Component Analysis (PCA) has been a core operation in data/signal processing \cite{Pearson,SVD_ref}.
Conceptually, PCA can be viewed as the pursuit of a coordinate system (defined by the principal components) that reveals underlying linear trends of a data matrix.
In its conventional form, the new coordinate system is calculated such that it preserves the energy content of the data matrix to the maximum possible extend.
Conventionally, the energy content of a data point is expressed by means of its L2-norm --i.e., its Euclidean distance from the center of the coordinate system. 
Thus, for any complex data matrix $\m X = [\m x_1, \m x_2, \ldots, \m x_N] \in\mathbb C^{D\times N}$, PCA searches for the size-$K$ ($1\leq K<\text{rank}(\m X)$) orthonormal basis (or, $K$-dimensional coordinate system) that solves
\begin{align}\label{L2Prob}
	\opt Q_{L2}=\argmax{\m Q\in\mathbb C^{D\times K};~\m Q^H\m Q=\m I_K}{\|\m Q^H\m X \|_2}
\end{align}
where, for any $\mathbf A \in \mathbb C^{m \times n}$, its L2-norm\footnote{The L2-norm of a matrix is also known as its  Frobenius or Euclidean norm \cite{Golub, Meyer}.} is defined as $\| \mathbf A \|_2 = \sqrt{\sum_{i=1}^m \sum_{j=1}^n |A_{i,j}|^2}$,  $| \cdot |$ denotes the magnitude of  a complex number (coinciding with the absolute value of a real number), and $\m I_K$ is the size-$K$ identity matrix. Due to its definition in \eqref{L2Prob}, PCA is also commonly referred to as L2-norm PCA, or simply L2-PCA.

A  practical reason of the tremendous popularity of L2-PCA is the computational simplicity by which the solution to \eqref{L2Prob} can be obtained. Specifically, a solution matrix $\opt Q_{L2}$ can be formed by the $K$ dominant singular vectors of $\m X$ and is, thus, obtainable by means of Singular-Value Decomposition (SVD) of $\m X$, with  quadratic complexity in the number of data samples $N$ \cite{Golub}.
Moreover, L2-PCA is a scalable operation in the sense that the $(k+1)$-th PC can be calculated using directly the first $k$ PCs that are always preserved.
In addition, there are several algorithms that can efficiently update the solution to \eqref{L2Prob} as new data points become available \cite{streamPCA}.
Finally, by the Projection Theorem \cite{Golub}  it is easy to show  that the maximum-L2-norm-projection problem in \eqref{L2Prob} is equivalent to the familiar minimum-L2-norm-error problem 
\begin{align}\label{L2minErr}
	\underset{\substack{\m Q\in\mathbb C^{D\times K};~ \m Q^H\m Q=\m I_K \\ \m Z \in\mathbb C^{K\times N}}}{\min.}{\|\m X-\m Q\m Z\|_2}.
\end{align}

On the downside, conventional L2-PCA,  seeking to maximize the L2-norm of the projected data-points in \eqref{L2Prob},  is well-known to be overly sensitive to outlying measurements in the processed matrix.
Such outliers may leek into the data matrix due to a number of different causes, such as  sensing/hardware malfunctions, external interference, and errors in data storage or transcription. 
Regardless of their cause, outliers are described as unexpected, erroneous values that lie far from the nominal data subspace and affect tremendously a number of data analysis methods, including L2-PCA.
Since the original conception of L2-PCA \cite{Pearson}, engineers and mathematicians have been trying robustify its against outliers. 
Popular robust versions of PCA are weighted PCA (WPCA) \cite{WPCA,rank1WPCA}, influence-function PCA \cite{pFunctions}, and L1-norm PCA (or, simply L1-PCA) \cite{L1PCA_0, L1PCA_m1,L1PCA_1, L1PCA_1v2, L1PCA_1v3, L1PCA_2,L1PCA_4,L1PCA_6,L1PCA_9,L1PCA_10,L1PCA_11,L1PCA_12,L1PCA_14,L1PCA_18,L1PCA_19,L1PCA_20,L1PCA_22,L1PCA_25,L1PCA_26,L1PCA_27,L1PCA_28, L1PCA_100, L1PCA_101, L1PCA_102, L1PCA_103, L1PCA_104, L1PCA_105, L1PCA_106, L1PCA_107, L1PCA_108}.

From an algebraic viewpoint, of all robust versions of PCA,  L1-PCA is arguably the most straightforward modification.
Mathematically, L1-PCA of real-valued data is formulated as
\begin{align} \label{L1ProbR}
\underset{\m Q\in\mathbb R^{D\times K};~\m Q^T\m Q=\m I_K}{\max.}{ \|\m Q^T\m X \|_1}
\end{align}
where $\| \cdot \|_1$ is the L1-norm operator, such that for any $\mathbf A \in \mathbb C^{m \times n}$, $\| \mathbf A \|_1 = \sum_{i=1}^m \sum_{j=1}^n |A_{i,j}|$. That is, L1-PCA derives from L2-PCA, by substituting the L2-norm with the more robust L1-norm. By not  placing squared emphasis on the magnitude of each point (as L2-PCA does), L1-PCA is far more resistant to outlying, peripheral points. 
Importantly, thorough recent studies have shown that when the processed data are not outlier corrupted, then the solutions of L1-PCA and L2-PCA describe an almost identical subspace. 

Due to its outlier resistance, L1-PCA of real-valued data matrices has attracted increased documented research interest in the past decade. 
Interestingly, it was  shown that real-valued L1-PCA can be converted into a combinatorial problem over antipodal binary variables ($\pm 1$), solvable with intrinsic complexity polynomial in the data record size,  $\mathcal{O}(N^{\text{rank}(\m X)K-K+1})$ \cite{L1PCA_m1}.

Despite its increasing popularity for outlier-resistant processing real-valued data, L1-PCA for complex-data processing remains to date unexplored. Similar to \eqref{L1ProbR},  complex L1-PCA is formulated as 
\begin{align} \label{L1Prob}
	\opt Q_{L1}=\argmax{\m Q\in\mathbb C^{D\times K};~\m Q^H\m Q=\m I_K}{ \|\m Q^H\m X\|_1}.
\end{align}
Interestingly, in contrast to real-valued L1-PCA, complex L1-PCA in \eqref{L1Prob} has no obvious connection to a combinatorial problem. 
Moreover, no finite-step algorithm (exponential or otherwise) has ever been reported for optimally solving \eqref{L1Prob}. 
Yet, as a robust analogous to complex L2-PCA, complex L1-PCA in  \eqref{L1Prob} can be traced to many important applications that  involve complex-valued measurements, e.g.,   in the fields communications, radar processing, or general signal processing, tailored to  complex-domain transformations (such as Fourier)  of real-valued data.

Our contributions in this present paper are summarized as follows.
\begin{enumerate}
\item We prove that \eqref{L1Prob} can be cast as an optimization problem over the set of unimodular matrices.\footnote{In this work a matrix is called unimodular if every entry has values on unitary complex circle. A unimodular matrix under our definition is not to be confused with the integer matrices with $\{-1,0,+1\}$-ternary minors.}
\item We provide the first two fast algorithms to solve \eqref{L1Prob} suboptimally.
\item We offer numerical studies that evaluate the performance of our complex L1-PCA algorithms. 
\end{enumerate}
Importantly, our  numerical studies  illustrate that the proposed complex L1-PCA exhibits sturdy resistance against outliers, while it performs similarly to L2-PCA when the processed data are outlier-free. 

The rest of the paper is organized as follows.
Section II offers as brief overview of technical preliminaries and notation.
Section III is devoted to the presentation of out theoretical findings and the derivation of the proposed algorithms.
Section IV holds our numerical studies.
Finally, some concluding remarks are drawn in Section V.

\section{Preliminaries and Notation}
Our subsequent algebraic developments involve extensively the \emph{sign} of a complex number and the \emph{nuclear norm} of a matrix. In this  section, we provide the reader with the definitions of these two measures, as well as useful pertinent properties. 

\subsection{The Sign of a Complex Number}

\begin{figure}
	\centering
	\includegraphics[scale=1]{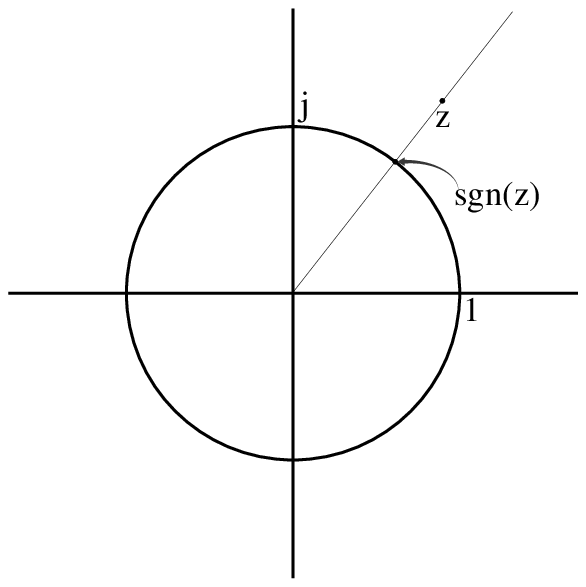}
	\caption{Graphical illustration of the sign of a complex number.}
	\label{Fig_sgn}
\end{figure}

Every complex number $z$ can be written as the product of its magnitude and a complex exponential. 
The complex exponential part is what we call ``sign" of the complex number and is denoted by  $\s{z}$. That is,  $\forall z\in\mathbb C: ~z = |z| \s{z}$, where $\s{z} \triangleq e^{j \angle z}$.
Clearly, the sign of any complex number belongs to the unitary complex circle 
\begin{align}
U\triangleq \{z\in\mathbb C:~ |z|=1\}.
\end{align}
 Fig. \ref{Fig_sgn} shows that the sign of any non-zero complex number is unique and satisfies the property presented Lemma \ref{signdist}.
\begin{mylemma}
For every $z \in C$, with $|z|>0$, $\s z$ is the point on $U$ that lies nearest to $z$ in the magnitude sense. That is, $\s z = \argmin{a\in U}{|a-z|}$.
\label{signdist}
\end{mylemma}
Through elementary algebraic manipulations, Lemma \ref{signdist} implies that 
\begin{align}\label{sgn_prop1}
	\s a = \argmax{b\in U}{\Re\{b^*a\}}.
\end{align}
In addition, the  optimal value of \eqref{sgn_prop1} is the magnitude of $a$. The above definition and properties of the sign can be generalized into vectors and matrices.
Let us define the sign of a matrix $\m A\in\mathbb C^{m\times n}$, for  any  $n$ and $m$, as the matrix that contains the  signs of the individual entries of $\mathbf A$. That is, we define
\begin{align}
	\s{\m A}\triangleq 
	\left[
	\begin{array}{ccc}
		\s{a_{1,1}} & \dots & \s{a_{1,n}} \\ \vdots & \ddots & \vdots \\ \s{a_{m,1}} & \dots & \s{a_{m,n}}
	\end{array} 
	\right].
\end{align}
In accordance to \eqref{sgn_prop1}, the sign of $\m A$ can be expressed as the solution to the maximization problem 
\begin{align} \label{sgn_prop1_mat}
\s{\m A}	 = \argmax{\m B\in U^{n\times m}}{ \Re\{\tr{\m B \m A} \}}.
\end{align}
Moreover, the optimal objective value of \eqref{sgn_prop1_mat} is the L1-norm of $\m A$; that is, 
\begin{align} \label{sgn_prop1_mat2}
\| \m A\|_1 = \underset{\m B\in U^{ n \times m}}{\max}~{ \Re\{\tr{\m B \m A} \}} = \tr{\text{sgn}(\m A)^H\m A}  .
\end{align}
Finally, by the above definitions it holds that  the sign of the product of two complex numbers equals the product of individual signs. In addition, it is clear that the sign of a number is $+1$ if-and-only-if the number is real and positive and $-1$ if-and-only-if the number is real and negative.

\subsection{The Nuclear Norm}
Consider matrix $\m A\in\mathbb C^{m\times n}$, with $m > n$ with no loss of generality. Then, let $\m A$ admit SVD $\mathbf A \overset{svd}{=} \m U  \text{Diag}(\boldsymbol \sigma) \m V^H$, where   $\m U^H \m U = \m V^H \m V = \m I_n$ and  $\boldsymbol \sigma \in \mathbb R_{\geq 0}^{n}$ contains the singular values of $\m A$ in descending order (i.e., $\sigma_{1} \geq \sigma_{2} \geq \ldots \geq \sigma_{n}$).\footnote{Consider  $\m a \in \mathbb C^{m}$ and $\m A=\text{Diag}(\m a)$; it holds $A_{i,i}=a_i$ for every $i \in \{1, 2, \ldots, m\}$ and $A_{i,j}=0$ for every $i \neq j$.} The nuclear norm of $\mathbf A$
is then defined as the summation of the singular values of $\m A$,
\begin{align}
	\|\m A\|_* \triangleq \sum_{i=1}^{r} \sigma_i = \| \boldsymbol \sigma \|_1.
\end{align}
Clearly, it holds that $\| \m A\|_* = \| \m A^H \|_*$. 
Being a fundamental quantity in linear algebra, the nuclear norm can be expressed in several different ways.
For example, in connection to the \emph{Orthogonal Procrustes Theorem} \cite{procrustes,Golub}, it holds  that 
\begin{align} \label{proc_label}
\| A \|_*=	\underset{\m Q\in\mathbb C^{m\times n};~\m Q^H\m Q=\m I_n}{\max}~{\Re\{\tr{\m Q^H\m A}\}}.
\end{align}
Moreover, denoting by $\text{unt}(\m A)$ the $m \times n$ unitary matrix that maximizes \eqref{proc_label} and assuming that  $\m A$ has full column rank (i.e., $\text{rank}(\m A)=n$), it holds that 
\begin{align}
\m A = \text{unt}(\m A)(\m A^H \m A)^\frac12,
\end{align}
which is known as the  \emph{polar decomposition} of $\m A$ \cite{polar,Golub}.
Finally,  $\text{unt}(\m A)$ can calculated by  the SVD of $\m A$ as 
\begin{align}
\unt{\m A} = \mathbf U \mathbf V^H.
\label{untSVD}
\end{align}

Based on the above preliminaries, in the following section we present our  developments on complex L1-PCA.

\section{Complex L1-PCA}
\subsection{Problem Connection to Unimodular Optimization}
In view of  \eqref{sgn_prop1_mat} and \eqref{proc_label} we can  rewrite the complex L1-PCA problem in \eqref{L1Prob} as
\begin{align}
	\underset{\m Q\in\mathbb C^{D\times K};~\m Q^H\m Q =\m I_K}{\max}~{\|\m Q^H \m X\|_1} & 
    =
    \underset{{\m Q\in\mathbb C^{D\times K};~\m Q^H\m Q=I_K}}{\max}~{ \tr{\text{sgn}(\m Q^H \m X)^H\m Q^H\m X} } \label{nuc_norm_opt1} \\
   & \overset{\eqref{sgn_prop1_mat}}= 
    \underset{\substack{\m Q\in\mathbb C^{D\times K};~\m Q^H\m Q=I_K
    \\ \m B\in U^{N\times K}}}{\max}~{\Re\left\{ \tr{\m B \m Q^H\m X} \right\}}  \label{nuc_norm_opt2}  \\
    & \overset{\eqref{proc_label} }= \underset{\m B\in U^{N\times K}}{\max}~{\|\m X\m B\|_*}. \label{nuc_norm_opt}
\end{align}

That is,  complex L1-PCA is directly connected to a maximization problem over the set of $N \times K$ unimodular matrices. 
Interestingly, Markopoulos et. al. \cite{L1PCA_m1,L1PCA_1} have proven a similar result for the real case. Specifically, \cite{L1PCA_m1} reformulated real-valued L1-PCA to a nuclear-norm maximization over  the set of $N \times K$ $(\pm 1)$-valued matrices,  $\{ \pm 1\}^{N \times K}$. Considering that   $\{ \pm 1\}$  is in fact the intersection of $U$ with the axis of real numbers,  we realize that the binary-nuclear-norm maximization to which real-valued L1-PCA corresponds \cite{L1PCA_m1} constitutes a relaxation of the unimodular-nuclear-norm maximization in \eqref{nuc_norm_opt}. Due to the finite size of $\{ \pm 1\}^{N \times K}$, finite-step algorithms could be devised for the solution of real-valued L1-PCA. Regretfully, since $U^{N \times K}$ has uncountably infinite elements, this is not the case for complex L1-PCA.

Even though unimodular-nuclear-norm maximization in \eqref{nuc_norm_opt} cannot be solved exhaustively,  there are still necessary optimality conditions that we can use to devise  efficient algorithms for solving it at least locally. The following proposition introduces the first of these optimality conditions.
\begin{myprop}\label{prop1}
Let $(\opt{Q_L1}, \opt{B})$ be an optimal solution pair for \eqref{nuc_norm_opt2}. Then, it holds that 
	\begin{align}
		\opt{Q_{L1}} = \unt{\m X\opt B} ~~\text{and}~~ 		\opt B=\s{\m X^H \opt{Q_{L1}} }.  \label{opt_condition_K_general}   
	\end{align}
    Moreover, $\opt{Q_{L1}}$ is a solution to \eqref{L1Prob} and $\opt B$ is a solution to \eqref{nuc_norm_opt}.   
\end{myprop}
Proposition \ref{prop1} derives directly from \eqref{sgn_prop1_mat}, \eqref{proc_label},  and the fact that both \eqref{L1Prob} and \eqref{nuc_norm_opt} are equivalent to \eqref{nuc_norm_opt2}. Most importantly, this proposition establishes  that, if $\opt B$ is a solution to \eqref{nuc_norm_opt}, then $\opt Q_{L1} = \unt{ \m X \opt B} $ is a solution to the L1-PCA in \eqref{L1Prob}. Thus, one can focus on solving  \eqref{nuc_norm_opt} and then use its solution to derive the L1-PCs by means of simple SVD (see the definition of $\unt{\cdot}$ in \eqref{untSVD}).
In addition, the two equations in \eqref{opt_condition_K_general}  can be combined into forming a new pair of necessary optimality conditions that concern the individual problems \eqref{L1Prob} and \eqref{nuc_norm_opt}. The new optimality conditions are presented in the following Corollary \ref{cor1}.
\begin{mycol} \label{cor1}
Let $\opt{Q_{L1}}$ be a solution to \eqref{L1Prob}; then, it holds that $\opt{Q_{L1}} = \unt{\m X\s{\m X^H \opt{Q_{L1}}}}$. Let $\opt B$ be a solution to \eqref{nuc_norm_opt}; then, it holds that $\opt B=\s{\m X^H \unt{\m X \opt B}}$.
\end{mycol}

\subsection{Complex L1-PCA when rank$(\m X) < D$}
Consider $\m X \in \mathbb C^{D \times N}$ with $r=\text{rank}(\m X) < D$.  $\m X$ admits thin SVD\footnote{In ``thin SVD" a matrix is written only in terms of its singular vectors that correspond to non-zero singular values.} $\m X \overset{svd}{=} \m U_x \m S_x \m V_x^H$, where $\m U_x^H \m U_x = \m V_x^H \m V_x = \m I_{r}$ and $\m S_x$ is the $r \times r$ diagonal matrix that  contains the non-zero singular values of $\m X$. In accordance to \eqref{prop2} above, to obtain the $K \leq  r$ L1-PCs of  $\m X$, $\opt Q_{L1}$,  we can work in two steps: (i) obtain the solution $\opt B$ to \eqref{nuc_norm_opt} and (ii) conduct SVD on $\m X \opt B$ and return $\opt Q_{L1} = \unt{\m X \opt B}$. Let us focus for a moment on the first step. We observe that\footnote{
For any square matrix $\m Z \in \mathbb C^{n \times n}$, $\sqrt{\m Z}$ is defined such that $\m Z =  \sqrt{\m Z} \sqrt{\m Z}$.
Also, for any $\m A \in \mathbb C^{m \times n}$, it holds that $\| \m A\|_* = \text{Tr}(\sqrt{\m A^H \m A})$. } 
\begin{align}
\| \m X \m B \|_* & =  \| \m U_x \m S_x \m V_x^H \m B\|_*  \label{short1}\\
& = \text{Tr} (\sqrt{ ( \m U_x \m S_x \m V_x^H \m B)^H\m U_x \m S_x \m V_x^H \m B }) \\
& = \text{Tr} (\sqrt{  \m B^H \m V_x^H \m S_x^H \m S_x \m V_x^H \m B }) \\
& = \| \m S_x \m V_x^H \m B\|_*  = \| \m X_{short} \m B \|_* \label{short2}
\end{align}
where $\mathbf X_{short} \triangleq \m S_x \m V_x^H \in \mathbb C^{r \times N}$. Then, $\opt B$ maximizes both \eqref{short1} (by definition) and \eqref{short2} (by equivalence). Notice also that  $\unt{\m X \opt B} = \m U_x \unt{\m X_{short} \opt B}$. By the above analysis, the following proposition holds true.  
\begin{myprop} 
Consider $\m X \in \mathbb C^{D \times N}$, with $r=\text{rank}(\m X) \leq \min \{D,N\}$, admitting thin SVD $\m X = \m U_x \m S_x \m V_x^H$ (i.e., $\m S_{x}$ is $r \times r$). Define $\m X_{short} = \m S_{X} \m V_x^H$. Let the $\opt Q_{L1, short} \in \mathbb C^{r \times K}$ be the $K < r$ L1-PCs of $\m X_{short}$, solution to $\underset{\m Q \in \mathbb C^{r \times K};~\m Q^H \m Q = \m I_{K}}{max.}~\|\m Q^H \m X_{short} \|_1$. Then, it holds that 
\begin{align}
\opt Q_{L1} = \m U_x \opt Q_{L1, short}
\end{align}
is the solution to the L1-PCA in \eqref{L1Prob}.  Moreover, $\| {\opt Q_{L1}}^H \m X\|_1 = \| \m X \opt B\|_* = \| \m X_{short} \opt B\|_* = \| {\opt Q_{L1, short}}^H \m X_{short} \|_1$.
\label{prop3}
 \end{myprop}
 Proposition \ref{prop3} shows that the L1-PCs of a rank-$r$ $D \times N$ matrix can always be obtained through the L1-PCA of a rank-$r$ $r \times N$ matrix. Therefore,  Proposition \ref{prop3}  steers our algorithmic focus to problems where $\m X$ has full row-rank (i.e., $D=r=\text{rank}(\m X)$).

\subsection{The Single-Component Case and L1-PCA Hardness}
In its simplest non-trivial form, complex L1-PCA is the search of a single ($K=1$) component  $\m q\in\mathbb C^{D\times 1}$ such that   $\|\m q^H X\|_1$ is maximized. 
In accordance to our more generic  developments for the multi-component ($K \geq 1$) case above,  the pursuit of a single L1-PC  can also be rewritten as a unimodular nuclear-norm maximization. That is, 
\begin{align}
\underset{\m q \in C^{D}; \| \m q\|_2=1}{\max}~\| \m q^H \mathbf X \|_1 &= \underset{\m b \in U^{N \times 1}}{\max}~\| \mathbf X \m b\|_* \label{l1pca1} \\
& = \underset{\m b \in U^{N \times 1}}{\max}~\| \mathbf X \m b\|_2 = \sqrt{\underset{\m b \in U^{N \times 1}}{\max}~ \m b^H \mathbf X^H \mathbf X \m b} \label{maxQuad}.
\end{align}
Equation in  \eqref{maxQuad} derives by the fact that any complex vector $\m a \in \mathbb C^{m}$ admits SVD $\m a \overset{svd}{=} \m u \sigma $, with $\m u=\mathbf a \| \m a\|_2^{-1}$ and $\sigma = \| \m a \|_2$ (trivially, the dominant right-hand singular vector is $1$).  By Proposition \ref{prop1}, for the L1-PC $\opt q_{L1}$ that solves \eqref{l1pca1} it holds 
\begin{align}
\opt q_{L1}  = \unt{\m X \opt b} = \m X \opt b \| \m X \opt b\|_2^{-1},
\label{b2q}
\end{align}
where $\opt b$ is the solution to \eqref{maxQuad}. Also, $\opt b = \s{\m X^H \opt q_{L1}}$.

The unimodular quadratic maximization (UQM) problem in \eqref{maxQuad} is a well-studied problem in the literature, with  several interesting applications, including among others the design of maximum-signal-to-interference-plus-noise-ratio (max-SINR) phased-arrays and and the design of unomodular codes \cite{Stoica}. 
For the real-data case, UQM in \eqref{maxQuad} takes the form of the  well- binary-quadratic-maximization \cite{Kary}, as proven in \cite{L1PCA_m1}.

Certainly, the necessary optimality condition presented in Corollary \ref{cor1} for \eqref{nuc_norm_opt} also applies to   \eqref{maxQuad}. Specifically, if $\opt b$ is a solution to \eqref{maxQuad}, then
\begin{align}
\opt b = \s{\m X^H \unt{\m X \opt b}} = \s{\m X^H \m X \opt b}.
\label{cond}
\end{align}
However, in contrast to what has been conjectured for the real case \cite{L1PCA_10,L1PCA_11} for the real case, \eqref{cond} is not a sufficient condition for local optimality. The reason is that $\m  b = \s{\m X^H \m X \m b}$ is also satisfied by some  ``saddle" points of $\| \m X \m b\|_2$. 
In this section we provide a  stronger optimality condition than \eqref{cond}, that is necessary and sufficient for local optimality. 

For compactness in notation, we begin by defining
\begin{align}
	\mom(\m b)  \triangleq\m b^*\odot (\m X^H \m X\m b) \in\mathbb C^{N\times 1},
\end{align}
 for any $\m b \in U^{N}$, where $(\cdot)^*$ performs complex-conjugation and $\odot$ is the element-wise product (Hadamard) operator. Even though the entries of $\mom(\m b)$ are complex, their summation $\sum_{n=1}^N \omega_n$ is real and positive, equal to the quadratic form $\m b^H \m X^H \m X \m b$. 
The following Proposition \ref{prop2} presents a necessary condition for optimality in the UQM of \eqref{maxQuad} that is satisfied by all local maximizers, but not by minimizers or saddle points.

\begin{myprop} \label{prop2}
	A unimodular vector ${\m b}$ is a local maximizer to \eqref{maxQuad},  if-and-only-if
	\begin{align}\label{key4}
		{\omega}_n(\m b)\in \mathbb R \text{~~and~~}  {\omega}_n(\m b) \geq \|\m x_n\|_2^2~~~\forall n\in \{1,  \ldots, N \}.
	\end{align}
\end{myprop}
\noindent \emph{Proof.} 
	For any  $\m b\in U^{N\times 1}$ there is an angle-vector $\mphi\in[0,2\pi)^{N\times 1}$  such that $\m b=e^{j\mphi}=[e^{j\phi_1},\dots,e^{j\phi_N}]^T$.
	The quadratic in the UQM of \eqref{maxQuad} can be then rewritten as $(e^{j\mphi})^H \m X^H \m X (e^{j\mphi})$, which is a function  continuously twice differentiable in the angle-vector $\mphi$; the corresponding first and second derivatives are
\begin{align}
		\m g(\m b) & =2\Im\{ \mom (\m b) \} \text{~~and} \label{grad} \\
 \m H(\m b) & =2\Re\Big\{\Diag (\m b)^H \m X^H\m X \Diag(\m b) -\Diag(\mom (\m b)) \Big\}, \label{hessian} 
\end{align}
	respectively.
	Any local maximizer of \eqref{maxQuad} will null the gradient and render the Hessian in \eqref{hessian} negative-semidefinite. 
	In the sequel we prove both directions of the equivalence (``if-and-only-if'' statement) in Proposition \ref{prop2}.
	
	\paragraph{Direct}
	If ${\m b}$ is local maximizer  of \eqref{maxQuad} then $\m H (\m b) \preceq 0$.
	Therefore, for every $n \in \{1, 2, \ldots, N\}$, 
	\begin{align}
		 \m e_n^T\m H  (\m b) \m e_n \leq 0  ~\Leftrightarrow~ 2\Re\Big\{  b_n^*\m x_n^H\m x_n   b_n - \omega_n(\m b) \Big\} \leq 0 ~  \Leftrightarrow ~ \omega_n(\m b) \geq \|\m x_n\|_2^2, 
	\end{align}
	where $\m e_n$ the $n$-th column of $\mathbf I_{N}$.
	
	\paragraph{Reverse}
	Consider ${\m b}$ such that all entries of ${\mom}(\m b)$  satisfy \eqref{key4}. 
	Then, for every $\mathbf z \in \mathbb C^{N}$, 
	\begin{align}  
    \m z^T\m H (\m b) \m z   
  & = 2 \m z^T \left(\Diag( {\m b} )^H \m X^H\m X\Diag ({\m b}) -\Diag( {\mom} (\m b))\right)  \m z  \\
  & = 2\left\| \sum_{n=1}^N z_n  b_n \m x_n  \right\|_2^2 - 2\sum_{n=1}^N z_n^2 \omega_n(\m b) \\
 	  & \leq 2\sum_{n=1}^N\left\|  z_n  b_n \m x_n  \right\|_2^2 - 2\sum_{n=1}^N z_n^2 \omega_n(\m b) \\
 & = 2\sum_{n=1}^Nz_n^2 \left( \left\| \m x_n  \right\|_2^2 - \omega_n(\m b) \right)  \leq 0
	\end{align}
	which implies that the Hessian at $\m b$,  $\m H (\m b)$, is negative-semidefinite. This,  in turn, implies that $\hat{\m b}$ is a local maximizer of \eqref{maxQuad}. \hfill $\square$
    
In the sequel, we present a direct Corollary of Proposition  \ref{prop2}.
\begin{mycol}
	A unimodular vector ${\m b}$ is a local maximizer to \eqref{maxQuad}, if-and-only-if
	\begin{align}\label{key5}
		{\m b}= \s{ \m A_d {\m b}}
	\end{align}
	where $\m A_d \triangleq \m X^H \m X - \text{Diag}([\| \m x_{1}\|_2^2, \ldots, \| \m x_{N}\|_2^2]^T)$.
\end{mycol}
\noindent \emph{Proof.} 
	In view of the sign of  the previous section, for every $n \in \{1, 2, \ldots, N\}$ it holds
	\begin{align}
		\omega_n(\m b) \geq \|\m x_n\|_2^2 & \Leftrightarrow \omega_n(\m b) -\|\m x_n\|_2^2 \geq 0  \\ 
		~ & \Leftrightarrow \s{\omega_n(\m b) -\|\m x_n\|_2^2}=1  \\ 
		~ & \Leftrightarrow \s{ b_n^* \m x^H\m X { \m b} -\|\m x_n\|_2^2}=1   \\
		~ & \Leftrightarrow  b_n= \s{ \sum_{m\neq n}\m x_n^H\m x_m  b_m }  	\end{align}
        which, by the definition of $\m A_d$, implies \eqref{key5}.
\hfill $\square$

The quantitative difference between the conditions \eqref{cond}   and \eqref{key5} lie  in the corresponding $\mom (\cdot)$ variables. 
On the one hand,  condition in \eqref{cond} guarantees that $\omega_n(\opt b)$ is positive; on the other hand, condition in \eqref{key5}  guarantees that, for every $n \in \{1, 2, \ldots, N\}$, $\omega_n(\opt b)$ is not only positive, but also greater-or-equal to $\|\m x_n\|_2^2$.
Hence,  \eqref{key5} is a clearly a stronger condition  than \eqref{cond}, in the sense that  \eqref{key5} implies  \eqref{cond} but not vise versa. For example, saddle points in $U^{N}$  could satisfy the mild condition in \eqref{cond} but not the necessary-and-sufficient local optimality condition in \eqref{key5}.

Proposition \ref{prop2} and  the corollary condition \eqref{key5} brought  us a step closer to solving \eqref{maxQuad} optimally. 
Specifically, based on \eqref{key5} we can also prove the following corollary.
\begin{mycol}\label{cor2}
	The UQM in \eqref{maxQuad} can be equivalently rewritten as
	\begin{align} \label{norm1Ad_prob}
		\underset{\m b\in U^{N\times 1}}{\max.}~{\|\m A_d\m b\|_1}.
	\end{align}
\end{mycol}
\noindent \emph{Proof.} 
	This problem equivalence results directly from \eqref{key5}.
	We have
	\begin{align*}
		\underset{\m b\in U^{N\times 1}}{\max}~{\m b^H\m X^H\m X\m b} & =  \underset{\m b\in U^{N\times 1}}{\max}~{\m b^H\m A_d\m b+\tr{\m X^H\m X}} \\
		~ & \overset{\eqref{key5}}{=} \underset{\m b\in U^{N\times 1};~  b=\s{\m A_d \m b}}{\max} {\m b^H\m A_d\m b+\tr{\m X^H\m X}} \\
		~ & =  \underset{\m b\in U^{N\times 1};~  b=\s{\m A_d \m b}}{\max}~{\|\m A_d\m b\|_1+\tr{\m X^H\m X}},
	\end{align*}
	which implies that  the UQM in \eqref{maxQuad} and the problemin \eqref{norm1Ad_prob} have identical optimal arguments and their optimal values differ by the constant $\tr{\m X^H\m X}$. \hfill $\square$ 
    
By Corollary \ref{cor2}, any effort to solve \eqref{norm1Ad_prob} counts toward solving the UQM and $K=1$ L1-PCA in \eqref{maxQuad}. 
Next, we discuss the hardness of UQM and complex L1-PCA ($K=1$). We notice that
UQM in \eqref{maxQuad} is, in fact, a quadratically-constrained quadratic program (QCQP) with concave objective function and non-convex constraints \cite{PardalosVavasis}.\footnote{In its standard form,  a QCQP is expressed as a minimization. Accordingly, the function that we minimize here is $-\m b^H \m X^H \m X \m b$, which is concave.} Therefore, it is formally  $\mathcal{NP}$-hard.
Since UQM in \eqref{maxQuad} is $\mathcal{NP}$-hard, the equivalent complex L1-PCA for $K=1$ is also $\mathcal{NP}$-hard. Accordingly complex L1-PCA in \eqref{L1Prob} must also be at least $\mathcal{NP}$-hard as a generalization of \eqref{maxQuad} for $K \geq 1$. 
In conclusion, in contrast to the real-field case of \cite{L1PCA_m1}, complex L1-PCA remains $\mathcal{NP}$-hard in the sample size $N$, even for fixed dimension $D$.

\subsection{Proposed Algorithms for Complex L1-PCA}

Based on the theoretical analysis above, in the sequel we present two algorithms for complex L1-PCA. Both algorithms are iterative and guaranteed to converge. 
With proper initialization, both algorithms could return upon convergence the global optimal solution of \eqref{nuc_norm_opt}. Our first algorithm relies on \eqref{cond} and can be applied for general $K$. Our second algorithm relies on the stronger condition \eqref{key5} and is applicable only to the $K=1$ case. 

\subsubsection{Algorithm 1}
For any given data matrix $\m X\in\mathbb C^{D\times N}$ and number of sought-after L1-PCs $K < \text{rank}(\m X)$, the algorithm  initializes at an arbitrary unimodular matrix $\m B^{(0)}\in U^{N\times K}$; then, in view of the mild optimality condition in \eqref{cond}, the algorithm   performs the iteration
\begin{align}
	\m B^{(i)}=\s{\m X^H\unt{\m X\m B^{(i-1)}}},~~i=1,2,\dots
    \label{algo1}
\end{align}
until the objective value in \eqref{nuc_norm_opt} converges. That is, the algorithm terminates at the first iteration $t$  that satisfies
\begin{align}\label{conv}
	\|\m X\m B^{(t)}\|_*-\|\m X\m B^{(t-1)}\|_* \leq  \delta
\end{align}
for some arbitrarily low convergence threshold $\delta \geq 0$. 
Then, the algorithm returns $\hat{\m B}$ as (approximate) solution to \eqref{nuc_norm_opt} and, in accordance to \eqref{opt_condition_K_general},  $\hat{\m Q}_{L1} = \unt{\m X\m B^{(t)}}$ as (approximate) solution to the L1-PCA problem in \eqref{L1Prob}. Below we provide a proof of convergence for the iterations in \eqref{algo1}, for any initialization $\m B^{(0)}$. 

\noindent \emph{Proof of Convergence of \eqref{algo1}.}
The convergence of \eqref{algo1} is guaranteed because the sequence $\{\|\m X\m B^{(i)}\|_*\}_{i=0}^{\infty}$ is (a) upper bounded by $\| \m X \opt B\|_*$ and (b) monotonically increasing; that is, for every $i$, 
 $\|\m X\m B^{(i-1)}\|_*\leq \|\m X\m B^{(i)}\|_* \leq \| \m X \opt B\|_*$. The monotonicity of the sequence can be proven as follows. 
	\begin{align}
		\|\m X\m B^{(i)}\|_* & = \max_{\m Q \in\mathbb C^{D\times K};~\m Q ^H\m Q = \m I_K}~{\Re\left\{ \tr{\m Q^H\m X\m B^{(i)}} \right\}} \label{aDummyLabel3} \\
		~ & \geq \Re\left\{ \tr{\unt{\m X\m B^{(i-1)}}}^H \m X\m B^{(i)}\right\} \label{aDummyLabel1}  \\
		~ & = \Re\left\{ \tr{\unt{\m X\m B^{(i-1)}}}^H \m X\s{\m X^H\unt{\m X\m B^{(i-1)}}}\right\}   \\
		~ &  =\max_{\m B\in U^{N\times K}}~\Re\left\{ \tr{\unt{\m X\m B^{(i-1)}}}^H \m X\m B'\right\} \label{aDummyLabel4} \\ 
		~ &  \geq \Re\left\{ \tr{\unt{\m X\m B^{(i-1)}}}^H \m X\m B^{(i-1)}\right\} \label{aDummyLabel2} \\ 
		~ & = \|\m X\m B^{(i-1)}\|_*.  
	\end{align}
	The inequality in \eqref{aDummyLabel1} holds because we have substituted $\m Q$ with a point in the the feasibility set of the maximization in \eqref{aDummyLabel3}, but not necessarily the maximizer. 
    	Similarly, inequality in \eqref{aDummyLabel2} holds because we have substituted $\m B$ with a point in the the feasibility set of the maximization in \eqref{aDummyLabel4}, but not necessarily the maximizer. 
\hfill $\square$

A detailed pseudocode of the proposed Algorithm 1 is presented in Fig. \ref{fig:algo1}. 

\subsubsection{Algorithm 2 (for $K=1$)}
Our second algorithm has the form of converging iterations, similarly to Algorithm 1. However, this algorithm relies on the strong optimality condition of \eqref{key5}, instead of the mild condition of \eqref{cond}. Specifically, given  $\m X\in\mathbb C^{D\times N}$ and an initialization point $\m b \in U^{N}$, Algorithm 2 iterates as
\begin{align}
	\m b^{(i)}=\s{\m A_d \m b^{(i-1)}},~~i=1,2,\dots
    \label{algo2}
\end{align}
until 
$
	\|\m A_d\m b^{(t)}\|_1-\|\m A_d\m b^{(t-1)}\|_1 \leq \delta
$
for some arbitrary small threshold $\delta$ and converging-iteration index $t$. Then the algorithm returns $\hat{\m b} = \m b^{(t)}$ as (approximate) solution to \eqref{maxQuad} and, in accordance to Proposition \ref{prop1} and \eqref{b2q},  $\hat{\m q}_{L1}={\m X \hat{\m b}}{\|\m X \hat{\m b} \|_2}^{-1}$ as approximate solution to \eqref{l1pca1}. Clearly, the iteration in \eqref{algo2} will converge because the sequence $\{\|\m A_d\m b^{(i)}\|_1\}_{i=1}^{\infty}$ is (a) upper bounded by $\| \m A_d \opt b\|_1$ and (b)  increases monotonically as 
	\begin{align}
		\|\m A_d \m b^{(i)}\|_1 & = \max_{\m b\in U^{N\times 1}}~\Re\left\{ \m b^H \m A_d \m b^{(i)}  \right\} \\
		~ & \geq \Re\left\{ {\m b^{(i-1)}}^H \m A_d \m b^{(i)} \right\} \\
		~ & = \Re\left\{ {\m b^{(i-1)}}^H \m A_d \s{\m A_d \m b^{(i-1)}} \right\} \\
		~ & = \|\m A_d \m b^{(i-1)}\|_1.
	\end{align}
A detailed pseudocode for Algorithm 2 is offered in Fig. \ref{fig:algo2}.

\section{Numerical Studies}

\subsection{Convergence}
\begin{figure}
	\centering
	\includegraphics[scale=1]{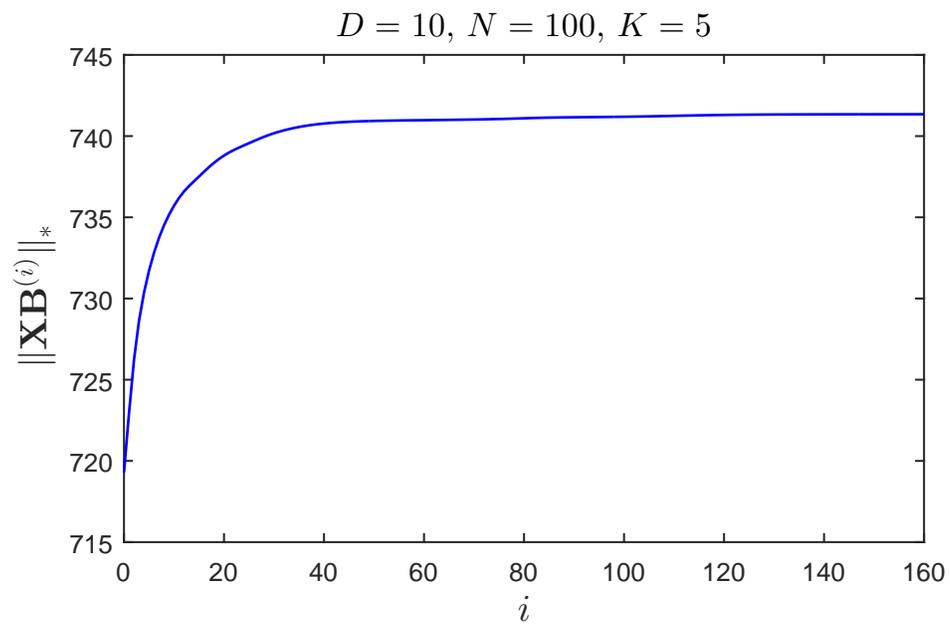}
	\caption{The objective metric of \eqref{nuc_norm_opt}, versus the iteration index $i$ for Algorithm 1 (single realization; $D=10$, $N=100$, $K=5$).}
	\label{fig:conv1}
\end{figure}

The convergence of Algorithm 1 was formally proven in the previous Section. At this point, to visualize the convergence, we fix $D=10$, $N=100$, and $K=5$, and  generate $\mathbf X \in \mathbb R^{D \times N}$ with entries drawn independently from $\mathcal {CN} (0, 1)$. Then, we run on $\m X$ Algorithm 1 (initialized at arbitrary $\mathbf B^{(0)}$) and plot in Fig. \ref{fig:conv1}   $\| \mathbf X \m B^{(i)}\|_*$ versus the iteration index $i$. We observe that, indeed, the objective nuclear-norm-maximization metric increases monotonically. 

\begin{figure}
	\centering
	\includegraphics[scale=0.7]{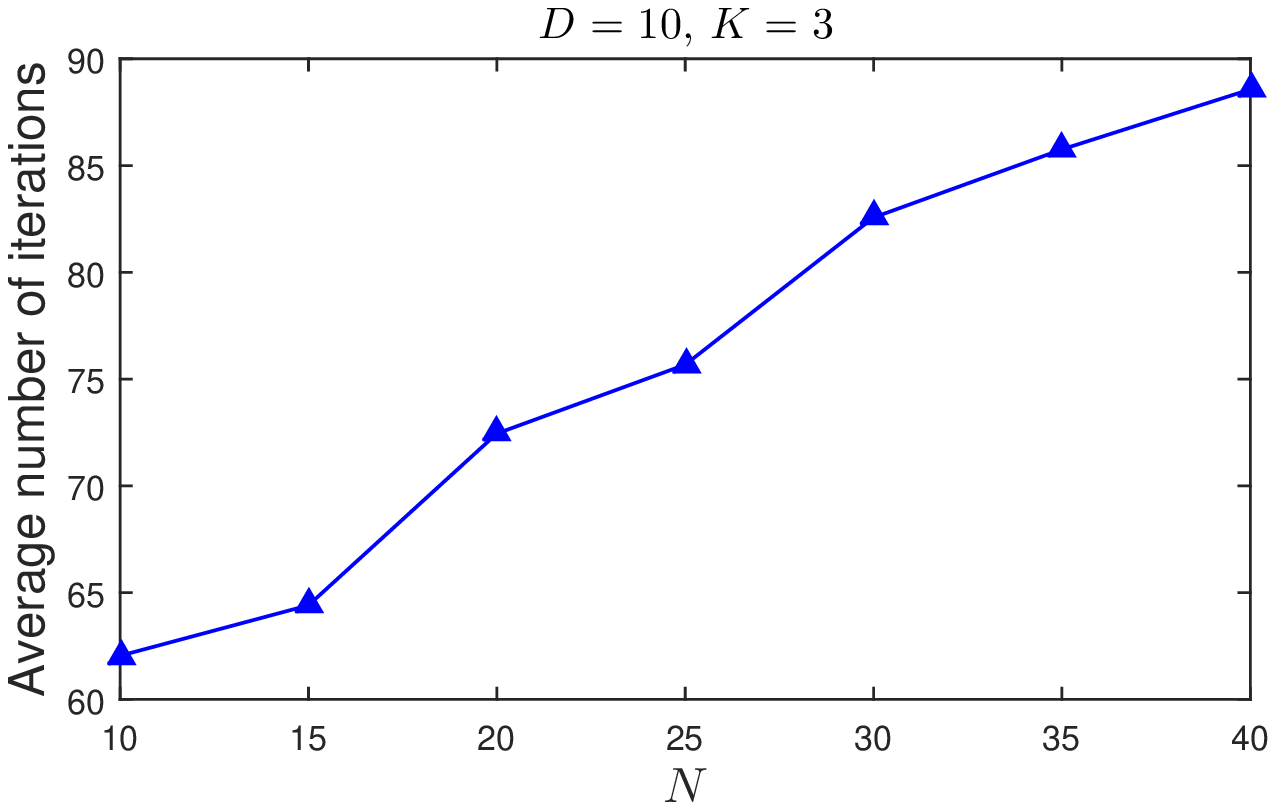}
	\caption{Average number of iterations needed for the convergence of Algorithm 1, versus the number of data points $N$ ($1000$ realizations; $D=10$, $K=3$).}
	\label{fig:vsN}
\end{figure}

\begin{figure}
	\centering
	\includegraphics[scale=0.7]{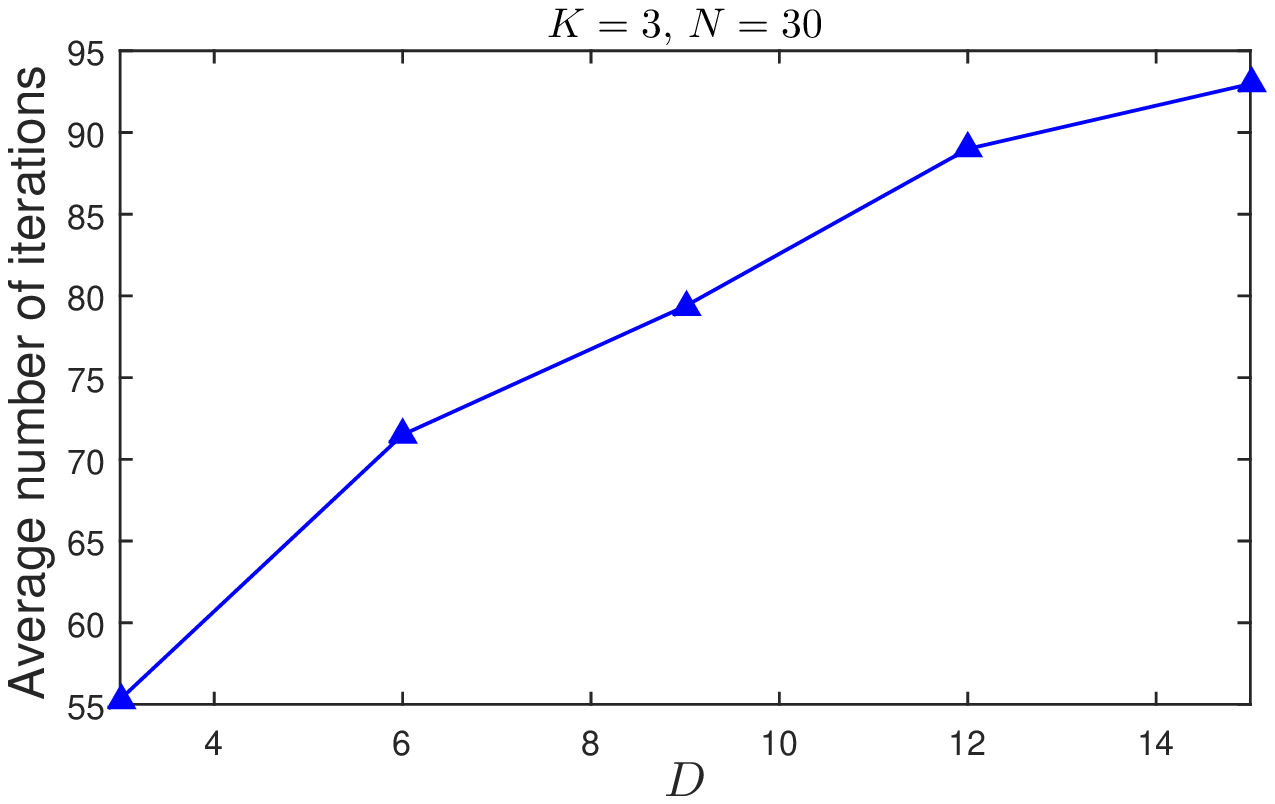}
	\caption{Average number of iterations needed for the convergence of Algorithm 1, versus the number of data points $N$ ($1000$ realizations; $K=3$, $N=30$).}
	\label{fig:vsD}
\end{figure}

\begin{figure}
	\centering
	\includegraphics[scale=0.7]{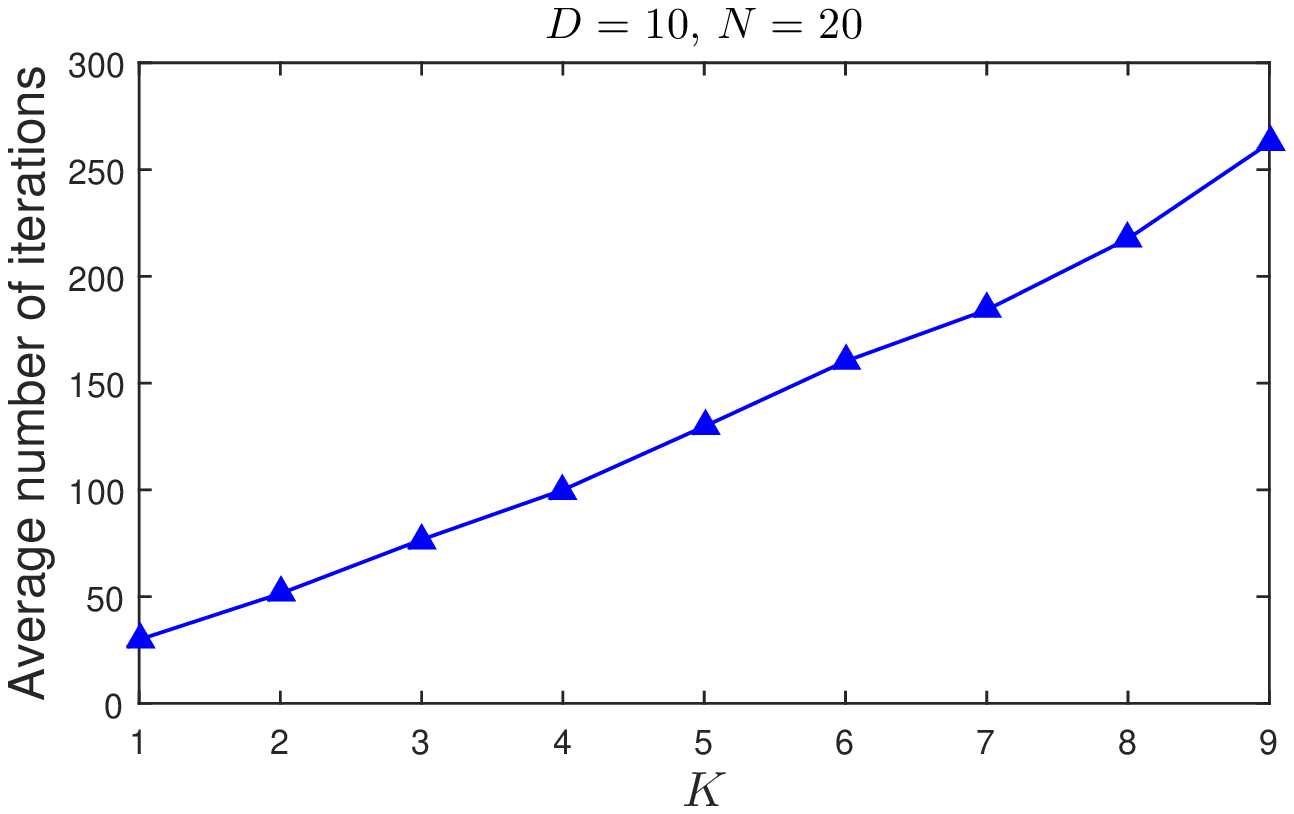}
	\caption{Average number of iterations needed for the convergence of Algorithm 1, versus the number of data points $N$ ($1000$ realizations; $D=10$, $N=20$).}
	\label{fig:vsK}
\end{figure}

Next, we wish to examine the number of iterations needed for Algorithm 1 to converge, especially as the problem-size parameters $D$, $N$, and $K$ take different values. 
First, we set $D=10$ and $K=3$ and vary $N =10, 15, \ldots, 40$. We draw again the entries of $\m X$ independently from $\mathcal{CN}(0,1)$ and plot in Fig. \ref{fig:vsN}  the average number of iterations needed for Algorithm 1 to converge (averaging is conducted over 1000 independent realizations of $\m X$) . We observe that, expectedly, the number of iterations increases along $N$. However, importantly, it appears to increase sub-linearly in $N$. In Fig. \ref{fig:vsD}  fix $K=3$ and $N=30$  and plot the average number of iterations needed for convergence versus $D$; in Fig. \ref{fig:vsK} we fix $D=10$ and $N=20$ and plot the average number of iterations versus $K$. We observe that the number of iterations increases sub-sub-linearly along $D$ and rather linearly along $K$. 

\subsection{Subspace Calculation}
In this first experiment, we  investigate and compare the outlier resistance of L2-PCA and L1-PCA. We consider the data matrix $\m X$ of \eqref{Xexp1}, consisting of $N=10$ data points of size $D=5$. A data processor wants to calculate the $(K=3)$-dimensional dominant subspace of $\m X$, spanned by its $K$ highest-singular-value left singular-vectors in $\mathbf Q_{n}  \in \mathbb C^{D \times K}$. However, unexpectedly,  1 out of the $N=10$ measurements (say, the first one) has been  additively corrupted by a random point $\m c $ drawn from $\mathcal{CN}(\m 0_{D}, \sigma^2 \m I_{D})$. Therefore, instead of $\mathbf X$, what is available  is the corrupted counterpart $\m X_{cor}= [ \m x_{1} + \m c, \m x_2, \ldots, \m x_N]$ (the data processor is not aware of the corruption). Instead of the nominal, sought-after $\text{span}(\m Q_{n})$, the data processor calculates the span of the $K$ L2-PCs of $\m X_{cor}$, $\m Q_{L2}$, and the span of the $K$ L1-PCs of $\m X_{cor}$, $\m Q_{L1}$. To quantify the corruption-resistance of the two subspace calculators, we measure the subspace proximity (SP)
\begin{align}
\text{SP}(\m Q, \m Q_{n}) = \frac{1}{\sqrt{K}}\| \m Q_{n}^H\m Q\|_2 \in [0,1],
\end{align}
for $\m Q = \m Q_{L2}$ and $\m Q=\m Q_{L1}$.
Certainly, if $\text{span}(\m Q)$ is orthogonal to the sought-after $\text{span}(\m Q_{n})$, then $\m Q^H \m Q_{n} = \m 0_{K \times K}$ and $\text{SP}(\m Q, \m Q_{n})=0$. On the other hand, if-and-only-if  $\text{span}(\m Q)$ coincides with $\text{span}(\m Q_{n})$, then $\text{SP}(\m Q, \m Q_{n})=1$. 

\begin{figure*}[t!]
\footnotesize{
\begin{align}
\mathbf X = 
\begin{bmatrix}
  -0.3003 - i~1.0117 &  0.4618 + i~0.0705 & -0.3924 - i~0.1602 & -0.5327 + i~0.1129 &  1.9368 - i~0.5685\\
  -0.3886 - i~0.6530 & -0.6204 - i~0.3556 &  0.7040 - i~1.3574 &  0.3315 + i~0.9675 & -1.5390 - i~0.8711\\
  -0.5961 + i~0.1708 &  0.6005 - i~1.8511 & -0.5541 - i~0.6086 & -0.4701 + i~0.3234 &  1.0896 + i~1.3071\\
  -0.0893 + i~0.1863 & -0.6031 + i~0.3869 & -0.7038 + i~0.0123 &  1.0782 + i~1.4440 &  0.9593 - i~0.9096\\
  -0.1678 + i~1.7097 &  0.5883 - i~0.7234 & -0.5185 - i~0.2924 & -0.3291 - i~1.7799 & -1.1252 - i~0.5569\\
   0.2485 + i~0.6433 & -1.3913 - i~1.7947 &  0.1189 + i~0.1334 &  0.0509 - i~0.1326 & -1.2163 + i~0.4921\\
   0.5302 - i~0.1632 & -0.9533 - i~0.3757 &  1.4074 - i~1.2147 & -0.4419 + i~0.8734 & -0.8092 - i~0.6724\\
   0.0428 + i~0.6675 & -1.1010 + i~0.6750 &  0.6385 - i~0.7620 &  0.4554 + i~0.5840 & -0.7863 + i~1.2148\\
  -1.3608 + i~0.5011 &  1.0467 - i~0.1282 &  0.5043 + i~0.1808 &  0.2366 - i~0.8010 &  0.0459 - i~0.3441\\
   0.5409 - i~0.7822 &  0.0075 - i~1.5285 &  1.4829 + i~0.9075 & -0.5216 - i~0.0030 &  0.8504 + i~0.8860\\
\end{bmatrix}^T.
\label{Xexp1}
\end{align}
}
\hrule
\end{figure*}

\begin{figure}
 \centering
		\includegraphics[scale=.8]{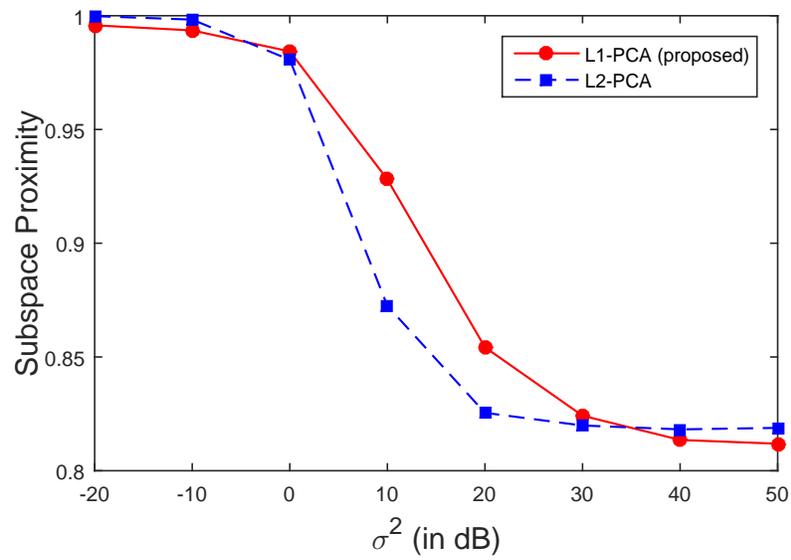}
		\caption{Subspace proximity versus corruption variance $\sigma^2$,  for L2-PCA and L1-PCA ($10~000$ realizations; $D=5$, $N=10$, $K=2$).}
        \label{spexp1}
\end{figure}

In Fig. \ref{spexp1} we plot the average SP (over $10~000$ independent corruption realizations) for L2-PCA and L1-PCA versus  the corruption variance $\sigma^2$. We observe that for weak corruption of variance  $\sigma^{2}<0$dB, L1-PCA and L2-PCA exhibit almost identical performance, with SP approaching the ideal value of $1$. We also notice that for very strong corruption of variance $\sigma^2 > 35$dB, both L1-PCA and L2-PCA get similarly misled converging to a minimum SP of about $0.82$. Interestingly, for all intermediate values of $\sigma^2$, L1-PCA exhibits significantly superior performance in calculating the nominal subspace. For example, for $\sigma^2=10$dB, L1-PCA attains $93\%$ SP, while L2-PCA attains $87\%$ SP. 

\subsection{Cognitive Signature Design}
Next, we investigate an application example for complex L1-PCA, drawn from the field of wireless communications. We consider a system of $K$ single-antenna primary sources  using unknown complex-valued spread-spectrum signatures of length $L$ chips.  The signatures of the $K$ sources are linearly independent (possibly, orthogonal), so that they do not interfere with each-other, spanning a $K$-dimensional subspace in $\mathbb C^{L}$. We consider now that $L-K$ secondary sources wish also to attempt using the channel, using length-$L$ spread spectrum signatures. 
Of course, the secondary sources should not interfere with the primary ones; for that reason, the $L-K$ signatures of the secondary sources should be orthogonal to the $K$ signatures of the primary sources --i.e., the secondary sources should transmit in the nullspace of the primary ones.  Therefore, the secondary users wish to estimate the subspace spanned by the primary signatures and then design signatures in its orthogonal complement. 

\subsubsection{Training Phase} 
With this goal in mind, we consider a collection of $N$ snapshots that correspond to primary transmissions in the presence of additive white Gaussian noise (AWGN); these snapshots will be used for estimating the primary-source signature subspace (and then design secondary signatures in its orthogonal complement). To make the problem more challenging, we consider that while these snapshots are collected, an  unexpected, strong  interference source is also sporadically active. 
That is, the $n$-th recorded snapshot vector (after down-conversion and pulse-matching), $\m x (n)\in \mathbb C^{L}$,  has the form
\begin{align} \label{model1}
\m x (n) = \underbrace{\sum_{k=1}^K \m s_k y_k(n) + \m n (n)}_{\text{nominal}} +  \underbrace{\gamma (n)  \m i(n)}_{
	\begin{smallmatrix}
	\text{unexpected} \\
	\text{corruption}
	\end{smallmatrix}
} \in\mathbb C^{L\times 1}.
\end{align}
In \eqref{model1},  $y_k(n)$ accounts for the product of the  $n$-th power-scaled information symbol transmitted by the $k$-th primary source with the flat-fading channel between the $k$-th source and the receiver, with $E\{|y_k(n)|^2\}=1$; $\m s_k \in \mathbb C^{L}$ is the signature of the $k$-th primary source, designed such that $\| \m s_{k}\|_2=1$ and $\m s_{k}^H \m s_{l}= 0$, for $k \neq l$; $\m n(n)$ is additive white Gaussian noise (AWGN), drawn from $\mathcal{CN}\left(\m 0_{L\times 1},\frac{1}{L}\m I_L\right)$; and $\m i(n)$ accounts for unexpected sporadic interference, drawn from $\mathcal{CN}\left(\m 0_{L\times 1},\frac{100}{L}\m I_L\right)$. $\{\gamma (n)\}_{n=1}^N$ are independent and identically distributed (i.i.d.)  $\{0,1\}$-Bernoulli($\epsilon$) variables that indicate interference activity. That is, each snapshot is corrupted by an unexpected interference signal with probability $\epsilon$.
According to the chosen values of symbol and noise variance, the primary users operate at signal-to-noise ratio (SNR) of $0$dB.
The recorded  snapshots are organized in the complex data record
$
\m X \triangleq [ \m x(1), \m x(2), \dots  \m x (N) ] \in\mathbb C^{L\times N}.
$
Then, we  analyze  $\m X$ to estimate the $K$-dimensional primary-source transmission subspace, $\mathcal S \triangleq \mathrm{span}( [\m s_{1}, $ $ \m s_{2}, \ldots, \m s_{K}] )$. Traditionally, $\mathcal S$ would be estimated as the span of 
$
\m Q_{L2} = \underset{\m Q \in \mathbb C^{L\times K}; ~\m Q^H \m Q=\m I_K}{\text{argmax}}{\| \m X^H \m Q \|_2}.
$
The reason is that, if all snapshots are nominal (i.e., no unexpected impulsive interference), as $N$ tends to infinity the span of $\m Q_{L2}$ provably coincides with that of the $K$ dominant eigenvectors of the snapshot autocorrelation matrix $E \{ \m x(n) \m x(n)^H\}$, which, in turn, coincides with $\mathcal S$. 
To examine the performance of complex L1-PCA, in this experiment  we also estimate $\mathcal S$ by the span of 
$
\m Q_{L1} = \underset{\m Q \in \mathbb C^{L\times K}; ~\m Q^H \m Q=\m I_K}{\text{argmax}}{\| \m X^H \m Q \|_1}.
$
After $\mathcal S$ is estimated, we pick $L-K$ orthogonal secondary signatures from the orthogonal complement of $\text{span}(\m Q_{L2})$ (or $\text{span}(\m Q_{L1})$). The secondary users employ these signatures and conduct transmissions concurrently with the primary sources. 

\subsubsection{Concurrent Operation of Primary and Secondary Sources} 
At this point, the assume that the impulsive corruption that interfered  with secondary-signature training is no longer active. As both the primary and secondary sources transmit, the receiver applies matched filtering for each primary source.  To evaluate the ability of L2/L1-PCA to identify the actual primary-source subspace (and thus enable secondary signatures that do not interfere with the primary sources), we measure and plot the post-filtering signal-to-interference-plus-noise ratio (SINR) for the primary users.
It is expected that if $\text{span}(\m Q_{L2})$ (or $\text{span}(\m Q_{L1})$) is close to $\mathcal S$, then the interference of the secondary-sources to the primary sources will be minimized and, accordingly, the aggregate post-matched-filtering SINR (sum-SINR) of all $K$ primary sources will be high.  We denote by $\m s_{l}'$ the signature of the $k$-th secondary source. It holds $\| \m s_{l}'\|_2=1$ and $\m s_{l}'^H\m s_{k}=0$ for $k \neq l$. Also, we denote by $y_{k}'(n)$ the $n$-th  symbol/channel compound for the $k$-th secondary source, with $E \{ | y_{k}'(n) |^2\}=\rho^2$ for all $k$. 
Sum-SINR is formally defined as $\text{sum-SINR}=\sum_{k=1}^K \text{SINR}_k$, where 
\begin{align} 
	\text{SINR}_k &  \triangleq \frac{ E\left\{ \left|\m s_k^H\m s_ky_k(n)\right|^2 \right\} }{E\left\{\left|\m s_k^H\left(  \sum_{m\neq k}\m s_m y_m(n)  +  \sum_{l=1}^{L-K} \m s_l'y_k'(n) + \m n (n)  \right)\right|^2\right\}}   \\
	~ & = \frac{1}{1+ \rho^2 \left|\sum_{l=1}^{L-K} \m s_{k}^H\m s_l'  \right|^2 }. \label{aDummyLabel5}
\end{align}
Certainly, sum-SINR is a decreasing function of the transmission energy  of the secondary sources, $\rho^2$.
We also observe that if $\mathcal S$ was perfectly estimated and, accordingly $\{ \m s_{k}'\}_{k=1}^{L-K}$ were designed in its orthogonal complement, then  $\m s_k^H\m s_l'=0$ for all $k,l$ and  \eqref{aDummyLabel5} takes its maximum value $1$ (or $0$dB),  independently of $\rho$. 

\begin{figure}
	\centering
	\includegraphics[scale=1]{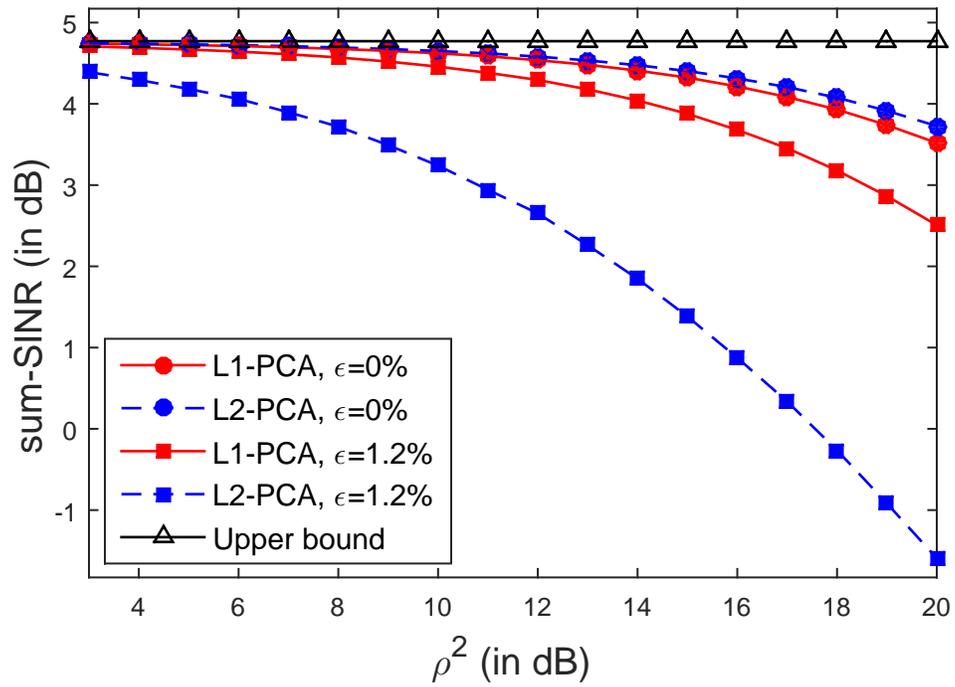}
	\caption{Post-filtering sum-SINR versus the transmission energy of the secondary sources, $\rho^2$ ($10~000$ realizations; $L=8$, $K=3$, $N=200$).}
	\label{cogSigDes_fig}
\end{figure}

In our numerical simulation we set $N=200$,  $L=8$, and $K=3$.
In Fig. \ref{cogSigDes_fig} we plot the average value of sum-SINR (calculated over $10~000$ independent experiments) versus $\rho^2$,  for snapshot-corruption probability $\epsilon = 0\%$ and $\epsilon = 1.2\%$.
As a benchmark, we also plot the horizontal line of  $10\log_{10}(K)$ dB, which corresponds to the sum-SINR if $\mathcal S$ was accurately estimated (i.e., $\m s_k^H\m s_l'=0$ for all $k,l$).
We observe that if $\epsilon=0$ (i.e., all $N$ training snapshots nominal),  L2-PCA-based and L1-PCA-based signature designs yield almost identical high sum-SINR performance. When however $\epsilon$ increases from $0$ to $1.2\%$, then L2-PCA  gets significantly impacted; accordingly, the sum-SINR performance of the L2-PCA-based signature design diminishes significantly. On the other hand, we observe that the L1-PCA-based design exhibits sturdy robustness against the corruption of the training snapshots, maintaining high sum-SINR performance close to the nominal one.

\subsection{Direction-of-Arrival Estimation}
Direction-of-Arrival (DoA) estimation is a key operation in many  applications, such as wireless node localization and network topology estimation.  Super-resolution DoA estimation relies, traditionally,   on the L2-PCA of a collection of --as, e.g., in MUltiple-SIgnal Classification (MUSIC)  \cite{MUSIC}. 

In this numerical study,  we consider a receiver equipped with a uniform linear array (ULA) of  $D$ antenna elements which receives signals from $K$ sources of interest located at angles $\Theta=\{\theta_1, \theta_2, \ldots, \theta_K \}$ with respect to the broadside. 
The inter-element spacing of the array, $d$, is fixed at half the wavelength of the received signal and  the array response vector for a signal that arrives from angle $\phi \in [\frac{-\pi}{2}, \frac{\pi}{2})$ is 
\begin{align}
	\m s(\phi) \triangleq [1,  e^{-j\pi d \sin(\phi)}, \ldots, e^{-j (D-1)\pi d \sin(\phi)}]^T \in\mathbb C^{D\times 1}.
\end{align}
To make the problem more challenging, we assume that apart from the  $K$ sources of interest, there are also $J$ unexpected, sporadically interfering sources (jammers),  impinging on the ULA from angles $\Theta' = \{\theta_1', \theta_2', \ldots, \theta_J' \}$. 
Therefore, the $n$th snapshot at the receiver is of the form
\begin{align} \label{model2}
	\m x (n) = \underbrace{\sum_{k=1}^K \m s({\theta_k}) y_k(n) ~~+ \m n (n)}_{\text{nominal}} + \underbrace{\sum_{j=1}^J\gamma_{n,j}\m s({\theta_j}') y_j'(n)}_{\text{unexpected jamming}}  \in\mathbb C^{D\times 1}.
\end{align}
In \eqref{model2}, $y_k(n)$ and $y_k(n)'$ account for the compound symbol of the $k$-th source and $j$-th jammer, respectively;  $\gamma_{n,j}$ is a $\{0,1\}$-Bernoulli($\epsilon$) activity indicator for jammer $j$ at snapshot $n$; $\m n (n)$ is the AWGN component, drawn from $\mathcal{CN}(\mathbf 0_D, \sigma^2 \mathbf I_{D})$.

The popular MUSIC DoA estimation method collects all $N $ snapshots in $\m X = [\m x(1), \m x(2), \ldots, \m x(N)]$ and calculates the source-signal subspace by the span of 
$
\m Q_{L2} = \underset{\m Q \in \mathbb C^{D\times K}; ~\m Q^H \m Q=\m I_K}{\text{argmax}}{\| \m X^H \m Q \|_2}.
$
If all snapshots are nominal, as $N$ tends to infinity $\text{span}(\m Q_{L2})$  tends to coincide with $\text{span}([\m s(\theta_1), \ldots, \m s(\theta_K)])$ and allows for accurate DoA estimation, by identifying the $K$  peaks of the, so called, MUSIC spectrum
\begin{align}
P(\phi; \m Q_{L2}) = \| (\mathbf I_{D} - \m Q_{L2} \m Q_{L2}^H ) \m s(\phi) \|_2^{-1}. 
\end{align}
In this work, we also conduct DoA estimation by finding the peaks  of  the L1-PCA spectrum $P(\phi;  \m Q_{L1})$, where 
$
\m Q_{L1} = \underset{\m Q \in \mathbb C^{D\times K}; ~\m Q^H \m Q=\m I_K}{\text{argmax}}{\| \m X^H \m Q \|_1}.
$.

In this numerical study, we set $D=12$, $K=4$, $J=3$,  source DoAs  $\Theta = \{-40^\circ, -21^\circ, -7^\circ, 60^\circ\}$, and jammer DoAs $\Theta' = \{ 0^\circ, 20^\circ, 80^\circ \}$. The number of snapshots available for DoA estimation (i.e., the  columns of $\m X$), $N$,  vary from $10$ to $100$ with step $10$.  The $K$ sources of interest operate at SNR 0dB; when active, jammers  operate at the much higher SNR of  15dB.  
We conduct $10~000$ independent DoA estimation experiments  and  evaluate the average DoA estimation performance of the L2-PCA-based and L1-PCA-based methods by means of the standard root-mean-squared error (RMSE), defined as
\begin{align}
	\text{RMSE} \triangleq \sqrt{\frac{1}{10000} \sum_{m=1}^{10000} \sum_{k=1}^K | \theta_{k}-\hat{\theta}_k(m)   |^2}
\end{align}
where $\hat{\theta_k}(m)$ is the estimate of $\theta_{k}$ at the $m$-th experiment. 

\begin{figure} 
 \centering
		\includegraphics[scale=.8]{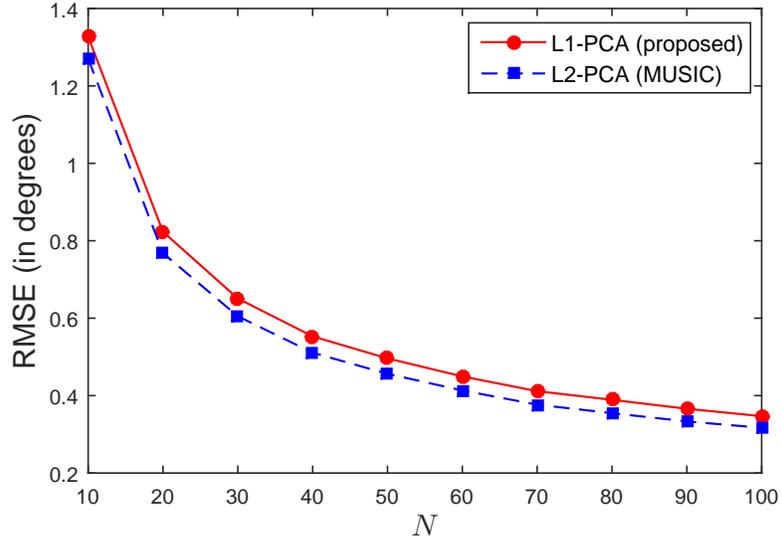}
		\caption{RMSE (in degrees) versus $N$, in nominal operation (no jammers; $\epsilon=0$), for L2-PCA (MUSIC method \cite{MUSIC}) and L1-PCA ($10~000$ realizations; $D=12$, $\Theta = \{-40^\circ, -21^\circ, -7^\circ, 60^\circ\}$, and $\Theta' = \{ 0^\circ, 20^\circ, 80^\circ \}$).}
		\label{NoJammer_Fig}
\end{figure}

\begin{figure}
 \centering
		\includegraphics[scale=.8]{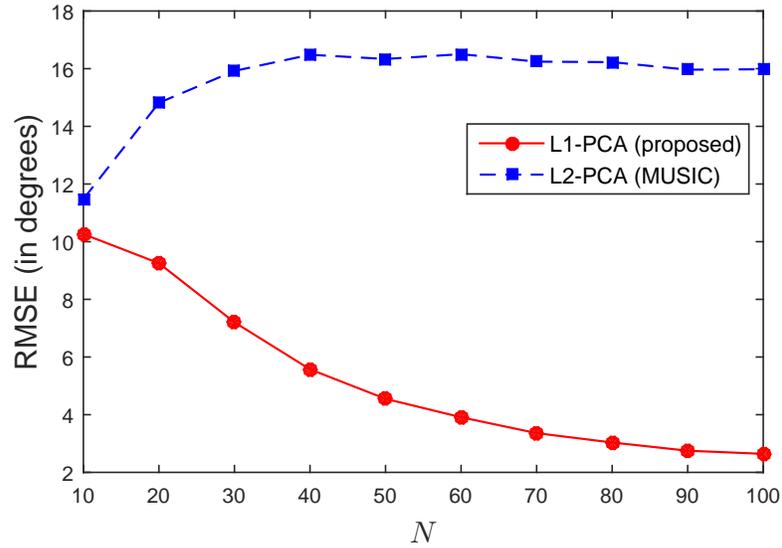}
		\caption{RMSE (in degrees) versus $N$, considering jammer corruption ($\epsilon=2\%$),  for L2-PCA (MUSIC method \cite{MUSIC}) and L1-PCA ($10~000$ realizations; $D=12$, $\Theta = \{-40^\circ, -21^\circ, -7^\circ, 60^\circ\}$, and $\Theta' = \{ 0^\circ, 20^\circ, 80^\circ \}$).}
				\label{Fig_with_jammer}
\end{figure}

In Fig. \ref{NoJammer_Fig} we set $\epsilon=0$ (i.e., no unexpected jamming) and plot the RMSE for L2-PCA and L1-PCA versus the number of snapshots, $N$. We observe that L2-PCA and L1-PCA have almost identical RMSE performance, which improves towards $0^\circ$ as $N$ increases.  Then, in Fig. \ref{Fig_with_jammer} we increase $\epsilon$ to $2\%$. 
The performance of L2-PCA (standard MUSIC \cite{MUSIC}) changes astoundingly. We observe that now the performance of MUSIC deteriorates as the sample-support $N$ increases converging on a plateau of poor performance at RMSE=$16^\circ$.
On the other hand, quite interestingly, L1-PCA resists the directional corruption of the jammers and, as $N$ increases it attains decreasing RMSE which reaches as low as $3^\circ$ for $N=100$. That is,  in contrast to L2-PCA, L1-PCA is benefited by an increased number of processed data points when the corruption  ratio remains constant.

\section{Conclusions}
We showed that, in contrast to the real-valued case, complex L1-PCA is formally $\mathcal{NP}$-hard in the number of
data points. Then, we showed how complex L1-PCA can be cast and solved through a unimodular nuclear-norm
maximization problem. We conducted optimality analysis and provided necessary conditions for global optimality.
For the case of K = 1 principal component, we provided necessary and sufficient conditions for local optimality. Based on the optimality conditions, we
presented the first two sub-optimal/iterative algorithms in the literature for L1-PCA. Finally, we presented extensive numerical studies
from the fields of data analysis and wireless communications and showed that when the processed complex data are
outlier-free, L1-PCA and L2-PCA perform very similarly. However, when the processed data are corrupted by faulty measurements
L1-PCA exhibits sturdy resistance against corruption and significantly robustifies  applications that rely on principal-component data feature extraction.

\newpage

\bibliographystyle{IEEEbib}

\newpage

\clearpage
 \newpage

\vfill 

\begin{figure}
	{\small
		{\hrule height 0.2mm} 
		\vspace{0.3mm}
		{\hrule height 0.2mm}
		\vspace{1mm}
		{\bf Algorithm 1: Iterative complex L1-PCA for general $K \geq 1$ (mild condition)}
		\vspace{0.5mm}
		{\hrule height 0.2mm}
		\vspace{2mm}
		\textbf{Input:} $\mathbf X \in \mathbb C^{D \times N}$;~ $K < \text{rank}(\m X)$;~ init. $\m B$; $\delta > 0$ \\
		\begin{tabular}{r l }
        	1: & $\alpha \leftarrow \| \m X \m B \|_*$\\
			3: & \texttt{while true}\\
			4: & $~~~~~$ $\m B \leftarrow \s{\m X^H \unt{\m X\m B}}$ \\
			5: & $~~~~~$ \texttt{if} $\| \m X \m B \|_* - \alpha > \delta$, $\alpha \leftarrow \| \m X \m B \|_*$ \\
            6: & $~~~~~$ \texttt{else, break} \\
            7: & $[\m U, \m S_{K \times K}, \m V] \leftarrow \text{svd}(\m X \m B)$\\
            8: & $\m Q \leftarrow \m U \m V^H$
		\end{tabular} \\
		\textbf{Output:} $\hat{\m Q}_{L1} \leftarrow \m Q$ and $\hat{\m B} \leftarrow \m B$
		\vspace{1.5mm}
		{\hrule height 0.2mm} 
        		\vspace{0.3mm}
		{\hrule height 0.2mm}
	}
	\caption{Proposed Algorithm 1 for the   L1-PCA of $\m X \in \mathbb C^{D \times N}$ (general $K \geq 1$). The algorithm relies on the mild optimality condition in \eqref{cond}.}
	\label{fig:algo1}
\end{figure}%

\begin{figure}
	{\small
		{\hrule height 0.2mm} 
		\vspace{0.3mm}
		{\hrule height 0.2mm}
		\vspace{1mm}
		{\bf Algorithm 2: Iterative complex L1-PCA for $K = 1$ (strong condition)}
		\vspace{0.5mm}
		{\hrule height 0.2mm}
		\vspace{2mm}
		\textbf{Input:} $\mathbf X \in \mathbb C^{D \times N}$;~ init. $\m b$; $\delta > 0$ \\
		\begin{tabular}{r l }
        	1: & $\m A_d= \m X^H \m X - \text{Diag}([\|\m x_{1}\|_2^2, \ldots,  \|\m x_{N}\|_2^2]^T)$, ~$\alpha \leftarrow \| \m A_d \m b \|_1$\\
			3: & \texttt{while true}\\
			4: & $~~~~~$ $\m b \leftarrow \s{\m A_d \m b}$ \\
			5: & $~~~~~$ \texttt{if} $\| \m A_d \m b \|_1 - \alpha > \delta$, $\alpha \leftarrow \| \m A_d \m b \|_1$ \\
            6: & $~~~~~$ \texttt{else, break} \\
            7: & $\m q \leftarrow \m X \m b \| \m X \m b\|_2^{-1}$
		\end{tabular} \\
		\textbf{Output:} $\hat{\m q}_{L1} \leftarrow \m q$ and $\hat{\m b} \leftarrow \m b$
		\vspace{1.5mm}
		{\hrule height 0.2mm} 
        		\vspace{0.3mm}
		{\hrule height 0.2mm}
	}
	\caption{Proposed Algorithm 2 for calculation of the $K=1$ L1-PC of $\m X \in \mathbb C^{D \times N}$. The algorithm relies on the strong optimality condition in \eqref{key5}.}
	\label{fig:algo2}
\end{figure}%

\vfill 

\clearpage
\newpage

\vfill 

\vfill 
\clearpage
\newpage

\vfill 

 \vfill 
\clearpage
\newpage
\vfill

\vfill 
\clearpage
\newpage
\vfill

 \vfill 
\clearpage
\newpage
\vfill
 
\end{document}